\documentclass[11pt]{article}
\usepackage[preprint]{acl}
\usepackage[T1]{fontenc}
\usepackage[utf8]{inputenc}
\usepackage{lmodern}
\usepackage{times}

\DeclareUnicodeCharacter{00B7}{\textperiodcentered}
\usepackage{latexsym}
\usepackage{microtype}
\usepackage{graphicx}
\usepackage{booktabs}
\usepackage{tabularx}
\usepackage{stfloats}
\usepackage{float}
\usepackage{multirow}
\usepackage{amsmath}
\usepackage{xcolor}
\definecolor{fidelitydiff}{HTML}{72AED3}
\newcommand{\fidelityscore}[2]{#1\textsubscript{\textcolor{fidelitydiff}{\normalfont\footnotesize #2}}}
\usepackage{colortbl}
\usepackage{hhline}
\usepackage{pifont}
\usepackage{tikz}
\usetikzlibrary{arrows.meta}

\newcommand{\keepwithfloat}[1]{\par\begingroup
  \dimen0=\pagegoal\advance\dimen0 by -\pagetotal
  \ifdim\dimen0<#1\relax\newpage\fi\endgroup}

\newcommand{\compacttable}{\small\setlength{\tabcolsep}{3pt}%
  \renewcommand{\arraystretch}{0.95}}
\newcommand{\fitwidth}[2]{\begingroup\sbox0{#2}%
  \ifdim\wd0>#1\resizebox{#1}{!}{\usebox0}\else\usebox0\fi\endgroup}

\newcommand{\sharedexampleline}[2]{%
  \noindent\hangindent=\dimexpr#1+1em\relax\hangafter=1\hspace*{#1}#2\par}
\newcommand{\sharedexamplecard}[2]{%
  \par\noindent\begingroup\setlength{\fboxsep}{7pt}\setlength{\fboxrule}{0.4pt}%
  \fcolorbox{black!20}{black!3}{%
    \begin{minipage}{\dimexpr\linewidth-2\fboxsep-2\fboxrule\relax}%
      {\sffamily\small\bfseries #1\par}\vspace{4pt}%
      {\color{black!18}\hrule}\vspace{6pt}%
      \normalfont\ttfamily\footnotesize\raggedright\setlength{\parskip}{0pt}%
      #2%
    \end{minipage}}%
  \endgroup\par\vspace{6pt}}

\title{LiveWorkspace: Externalizing Visual Understanding as Persistent Shared State for Screenshot-to-Code Agents}
\author{%
  {\normalsize\bfseries Shijun Xiao$^{1,2}$, Zonghui Yang$^3$, Xinrong Chen$^3$, Hao Liu$^2$, Wensheng Zhang$^4$, Yongqiang Tang$^{2,*}$}\\[5pt]
  {\small\normalfont $^1$Nankai University, Tianjin, China}\\[1pt]
  {\small\normalfont $^2$Institute of Automation, Chinese Academy of Sciences, Beijing, China}\\[1pt]
  {\small\normalfont $^3$Peking University, Beijing, China}\\[1pt]
  {\small\normalfont $^4$Guangzhou University, Guangzhou, China}\\[4pt]
  {\small\normalfont $^*$Corresponding author: \texttt{yongqiang.tang@ia.ac.cn}}%
}
\DeclareRobustCommand{\LWcompoundhyphen}{\leavevmode\nobreak\hskip0pt\relax-\discretionary{}{}{}\nobreak\hskip0pt\relax}
\hypersetup{pdfauthor={Shijun Xiao, Zonghui Yang, Xinrong Chen, Hao Liu, Wensheng Zhang, Yongqiang Tang},pdftitle={LiveWorkspace: Externalizing Visual Understanding as Persistent Shared State for Screenshot-to-Code Agents}}
\begin{document}
  \maketitle
  \begin{abstract}
Screenshot-to-Code requires coding agents to sustain visual understanding across code generation, execution, and revision. We introduce LiveWorkspace, a visual coding harness that externalizes this understanding as Shared Visual State (Shared VS), an implementation-independent representation of visual hierarchy, node attributes, and layout relations. LiveWorkspace constructs and retains \textit{Target} from screenshots and recovers \textit{Current} from browser execution in the same schema. Structured Diff and implementation bindings guide code modifications; user edits update \textit{Target} and reuse the alignment loop. Using GPT-5.2 for both methods, LiveWorkspace improves end-to-end Avg by 1.8 points on Design2Code and 2.0 on ScreenBench over a control with matched model-call counts and stagewise budget limits. Further experiments demonstrate reuse of constructed Shared VS across coders and improvements in iterative repair. LiveWorkspace also achieves 93.0\% edit success across 100 controlled editing cases. These findings support treating visual understanding as a persistent shared working state throughout code generation, execution, and modification.
\end{abstract}

  \section{Introduction}
\label{en:sec:introduction}

\begin{figure}[!t]
\centering
\includegraphics[width=\columnwidth]{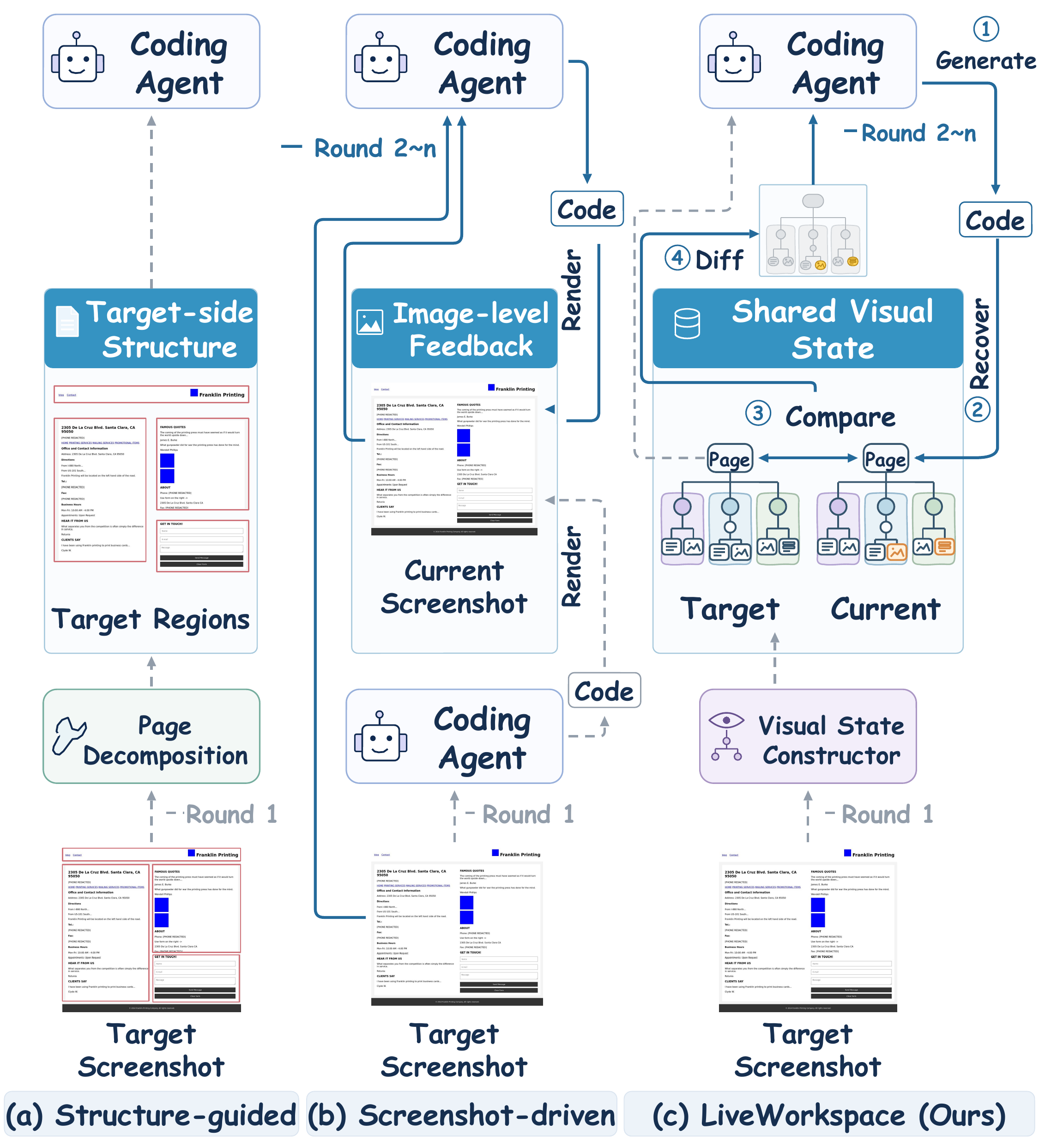}
\caption{Visual interfaces for Screenshot-to-Code. (a) Target-side structure guides generation; (b) rendered screenshots provide image-level feedback; (c) LiveWorkspace maintains Target and Current in Shared VS and exposes their differences for code revision.}
\label{en:fig:intro-overview}
\end{figure}

As coding agents evolve from code completion and program generation toward understanding,
building, and modifying real software interfaces \cite{sweagent,openhands,codeact}, Screenshot-to-Code is becoming a
fundamental capability for translating visual intent into executable software. Given a
target webpage screenshot, a model must generate HTML/CSS/JavaScript whose browser
rendering faithfully reproduces the target design. The model must identify visual elements, recover their structure and attributes, and
translate these visual constraints into executable code. Code generation is also not the end of this process: many visual deviations
become apparent only after the code is rendered, requiring the coding agent to verify
and revise the implementation. The central challenge of Screenshot-to-Code is therefore to sustain visual understanding that guides code generation and modification after execution.

Existing methods improve this process along two main directions. One line of work uses
explicit page decomposition and structural modeling to provide intermediate representations
of regions, layouts, or hierarchies for code generation, including DCGen, LaTCoder,
LayoutCoder, and ScreenCoder \cite{dcgen,latcoder,layoutcoder,screencoder}. Another brings
browser execution results back into generation and performs multiple rounds of refinement
based on differences between the target screenshot and the current rendering, including
UI2Code$^{N}$, VisRefiner, and other rendering-in-the-loop methods
\cite{ui2coden,visrefiner}. The former improves understanding of the target page structure;
the latter enables continued correction of visual errors through execution feedback after
initial generation. However, target-side structural representations and execution-side visual feedback still lack a common, persistent state interface. Structural representations primarily guide code generation, without normalizing execution results into the same representation. Screenshot-based feedback still relies on the model to infer visual correspondences between the target and current implementation from images.

Based on this observation, we formulate Screenshot-to-Code as a continuous process of
visual alignment and introduce \textbf{Shared Visual State (Shared VS)} to externalize page
understanding into a working state independent of the specific HTML/CSS implementation.
Shared VS describes the final visible structure and appearance of a page through visual
hierarchy $H$, node attributes $A$, and layout relations $L$, without requiring recovery
of the original DOM or CSS. The same schema supports recovery of \textit{Target} Shared VS
from the target screenshot and \textit{Current} Shared VS from code execution, enabling
explicit comparison between the target design and current implementation.

Building on this representation, we introduce \textbf{LiveWorkspace}, a visual coding
harness that uses Shared VS as its core working state and combines the coding agent's
structural reasoning with deterministic visual measurement. LiveWorkspace retains \textit{Target} Shared VS across iterations and recovers \textit{Current} Shared VS after each execution, using their differences to guide subsequent modifications; user edits update \textit{Target} and reuse the same alignment process. Figure~\ref{en:fig:intro-overview} illustrates the distinction between
target-side structure, image-level feedback, and persistent shared visual state.

We evaluate the complete LiveWorkspace harness on Design2Code and ScreenBench. Compared with a GPT-5.2 control with matched model-call counts and stagewise budget limits, LiveWorkspace improves final reconstruction Avg from $91.8$ to $93.6$ on Design2Code and from $81.2$ to $83.2$ on ScreenBench, gains of $1.8$ and $2.0$ points, respectively.
With only Recovered Target Shared VS as visual input, DeepSeek-V4 achieves initial-generation Avg scores of $92.7$ and $83.0$ on Design2Code and ScreenBench. Controlled repair experiments show that Structured Diff improves the repair rate, with further gains from source binding; LiveWorkspace also achieves higher final fidelity and introduces fewer new defects than screenshot feedback. Across $100$ controlled editing cases, LiveWorkspace achieves $93.0\%$ edit success.

Our main contributions are as follows:
\begingroup
\begin{itemize}
  \item We introduce Shared VS as an implementation\LWcompoundhyphen{}independent interface that represents
  target designs and executed implementations in a common visual schema for explicit
  comparison and alignment.
  \item We build a visual coding harness around Shared VS, separating agent-based structural
  reasoning from deterministic visual measurement. It retains \textit{Target}, recovers
  \textit{Current} after execution, and unifies generation, repair, and user editing through
  the same state interface.
  \item Our experiments show that Shared VS supports reuse across coding models, Structured
  Diff turns state discrepancies into actionable repair feedback, and source binding
  provides implementation-location cues that further improve repair.
\end{itemize}
\endgroup

  \section{Related Work}
\label{en:sec:related-work}

\subsection{Structured Visual Understanding for Screenshot-to-Code}
\label{en:sec:related-structure}

Screenshot-to-Code has progressed from early image-to-code generation to multimodal webpage generation and evaluation \cite{pix2code,websight,design2code}. Larger training datasets \cite{webcode2m,web2code}, broader benchmarks \cite{iwbench,designbench}, data synthesis \cite{flame}, and visual-reward training \cite{vincicoder} support this development. UI understanding integrates visual, textual, and structural information \cite{rico,uibert,screenai,ferretui2}, while screen parsing recovers elements and relations \cite{uied,screenparsing,omniparser,screenparse}. DCGen decomposes screenshots into regions, and UICopilot separates coarse hierarchy generation from fine-grained code synthesis \cite{dcgen,uicopilot}. LaTCoder uses Layout-as-Thought for block generation and assembly, LayoutCoder constructs element relations and layout trees \cite{latcoder,layoutcoder}, and ScreenCoder separates grounding, planning, and generation \cite{screencoder}. LiveWorkspace extends this structural perspective with a common, implementation-independent schema for target designs and executed implementations, enabling explicit visual-state comparison.

\subsection{Visual Feedback and Iterative UI Refinement}
\label{en:sec:related-feedback}

Self-Refine establishes a general feedback-and-refinement loop \cite{selfrefine}, while Sketch2Code studies interactive webpage prototyping with user feedback \cite{sketch2code}. For Screenshot-to-Code, UI2Code$^{N}$ uses rendered feedback for iterative visual optimization \cite{ui2coden}, and VisRefiner learns from visual differences and corresponding code edits \cite{visrefiner}. ComUICoder uses element-level matching to produce targeted instructions for adding or removing elements and correcting visual attributes \cite{comuicoder}. Visual software repair benchmarks screenshot-bearing issues \cite{swebenchmm} and explores cross-modal reasoning, region-focused perception, and rendered feedback \cite{guirepair,visualrepair,codingwitheyes}. Reflexion and FailureMem retain feedback or repair experience across attempts \cite{reflexion,failuremem}. LiveWorkspace retains the desired page state as \textit{Target} and recovers \textit{Current} after each execution. Their shared schema makes visual discrepancies explicit, while implementation bindings connect them to code locations. User edits update \textit{Target} and reuse this alignment loop.

  \begin{figure*}[!t]
\centering
\includegraphics[width=\textwidth]{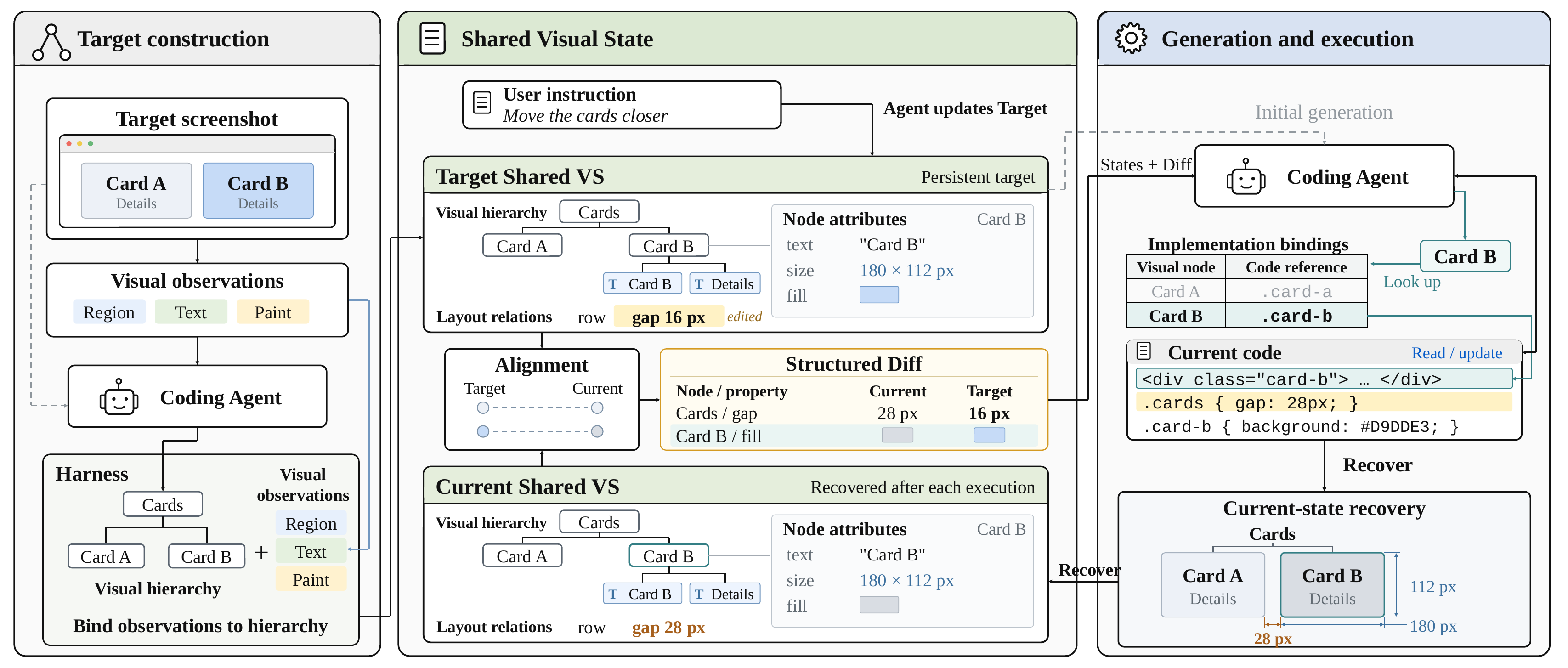}
\caption{Overview of LiveWorkspace. The agent uses the screenshot and harness-provided observations to infer visual hierarchy. The harness binds observations to this hierarchy and recovers attributes and layout, forming \textit{Target} Shared VS $S=(H,A,L)$. Initial generation uses only \textit{Target}. \textit{Current} is recovered from browser execution in the same schema. Structured Diff and implementation bindings guide code repair. \textit{Target} persists across iterations; user edits update it and reuse the alignment loop.}
\label{en:fig:svs-construction}
\end{figure*}

  \section{Method}
\label{en:sec:method}

As shown in Figure~\ref{en:fig:svs-construction}, LiveWorkspace is a Screenshot-to-Code harness whose core working state is
Shared Visual State (Shared VS). Given a target screenshot $I^t$ and code $C_k$ at iteration $k$, the system maintains the
\textit{Target} state $S^t$, \textit{Current} state $S^c_k$, structured difference $\Delta_k$,
and an independent implementation binding $B_k$.
The same coding agent performs page-structure reasoning and code decisions, while the
harness handles visual observation, state recovery, comparison, and execution feedback.
This agent--harness interaction follows the reasoning--action paradigm \cite{react}.
The overall process is
\begin{align}
  S^t &= \mathrm{Construct}(I^t), \\
  S^c_k &= \mathrm{Recover}(C_k,\mathrm{Render}(C_k)), \\
  \Delta_k &= \mathrm{Diff}(S^t,S^c_k), \\
  C_{k+1} &= \mathrm{Agent}(S^t,S^c_k,\Delta_k,B_k,C_k).
\end{align}
Initial generation uses only \textit{Target} Shared VS, which is retained across refinement
and updated for user edits. After each execution, the harness recovers \textit{Current} Shared
VS in the same schema, recomputes the structured difference, and refreshes the implementation
binding.

\subsection{Shared Visual State}
\label{en:sec:method-svs}

Shared VS describes the visible state of a page rather than the DOM or CSS
that produces it. Since different implementations can yield identical or similar renderings \cite{webrenderbench},
we canonicalize a page as
\begin{equation}
  S=(H,A,L),
\end{equation}
where $H$ denotes visual hierarchy, $A$ denotes node attributes, and $L$ denotes layout
relations between nodes.

\textbf{Visual hierarchy $H$.} $H$ describes visual grouping and containment, with internal
nodes for meaningful regions or groups and leaves for visible elements such as text and
images. Synthetic groups can express visual groupings without corresponding source-code
elements.

\textbf{Node attributes $A$.} $A$ describes observable visual facts about each node,
including position and size, text, foreground color, fill, and border.

\textbf{Layout relations $L$.} $L$ describes how elements within the same visual group are
arranged, including sequence, grid, and overlay, as well as recoverable direction, spacing,
and padding. $H$ describes which elements belong to the same group, while $L$ describes how elements
within that group are arranged.

During execution, LiveWorkspace maintains two instances of the same Shared VS
schema: Target Shared VS describes how the page should appear, and Current Shared VS
describes the visual state actually produced by the current code. DOM nodes, selectors, and source locations are stored in an independent
binding and are excluded from Shared VS itself.

\subsection{Target Shared VS Construction}
\label{en:sec:method-target}

\textit{Target} Shared VS construction separates agent-based structural reasoning from
deterministic visual measurement.

For the target screenshot $I^t$, the harness first extracts three types of low-level
visual observations:
\begin{equation}
  O=O_{\mathrm{region}}\cup O_{\mathrm{text}}\cup O_{\mathrm{paint}},
\end{equation}
which provide region candidates, text and its geometry, and paint evidence such as color,
fill, and border, respectively.

The coding agent uses the target screenshot and observations $O$ to infer visual hierarchy $H^t$, identifying visual units and their grouping and containment relations. Given $H^t$, the harness associates observations with canonical nodes: node attributes are recovered from the associated text, geometry, and paint evidence, while layout relations are inferred from the painted geometry of children within each group. The agent thus supplies page organization, and the harness fills the representation with visual facts supported by the observations, yielding the complete state $S^t=(H^t,A^t,L^t)$:
\begin{equation}
  S^t=\mathrm{Recover}(I^t,O,H^t).
\end{equation}

\subsection{Current Shared VS and Alignment}
\label{en:sec:method-current}

After code execution, LiveWorkspace recovers Current Shared VS from the current rendering:
\begin{equation}
  S^c_k=(H^c_k,A^c_k,L^c_k).
\end{equation}
Current and Target use the same canonical visual schema but rely on different recovery
evidence. Target uses the screenshot and visual observations, whereas Current uses the
runtime DOM, computed styles, and actual painted geometry. Through Visual-Tree Canonicalization, the harness collapses wrappers without visual meaning, prunes invisible elements, and infers layout relations from painted geometry. The DOM serves only as recovery evidence; implementation identifiers such as tags, classes, and selectors are excluded from $S^c_k$.

Since \textit{Target} and \textit{Current} are recovered independently and do not share node IDs,
the harness first establishes node correspondences. Nodes with visible text are matched
exactly using normalized text; multiple candidates with identical text are disambiguated
by geometric center distance. Remaining nodes are matched greedily, one-to-one, using
the IoU of absolute page bounding boxes ($IoU\geq0.5$), prioritizing pairs with the largest
overlap. Text matching does not require box overlap, so displacement is reported as an attribute difference for text-matched nodes.

The harness compares visual attributes and within-group layout for corresponding nodes, and reports unmatched reportable visual nodes as structural differences.
\begin{equation}
  \fitwidth{\dimexpr\columnwidth-3em\relax}{$\displaystyle
    \Delta(v)=\{\Delta_{\mathrm{geometry}},\Delta_{\mathrm{text}},
    \Delta_{\mathrm{color}},\Delta_{\mathrm{layout}},\ldots\}.$}
\end{equation}
Attribute differences record current and target values; numerical attributes use tolerances to filter minor measurement deviations.

Alongside \textit{Current} Shared VS, the harness maintains an implementation binding separate from Shared VS:
\begin{equation}
  \fitwidth{\dimexpr\columnwidth-3em\relax}{$\displaystyle
    B_k:v^c\leftrightarrow\{\text{DOM node},\text{selector},\text{source span}\}.$}
\end{equation}
For generated single-file HTML, the harness establishes implementation bindings using element types, text containment, and document order. Structured Diff identifies visual attributes to change and their current and target values; binding provides implementation-location cues for the corresponding \textit{Current} nodes. The agent combines this information with the current code to decide how to modify the implementation.

\subsection{Generation, Refinement, and Editing}
\label{en:sec:method-editing}

Initial code generation uses \textit{Target} Shared VS as visual input and produces implementation $C_0$. After execution, the harness recovers \textit{Current} and computes Diff and implementation bindings; the coding agent uses them to choose a local patch or a broader modification. The modified code is executed again to produce the next \textit{Current}, while \textit{Target} remains unchanged throughout refinement.

Given a user instruction $u$, the agent identifies the relevant visual nodes and updates the affected hierarchy, attributes, or layout relations in \textit{Target} Shared VS, yielding $S^{t\prime}$:
\begin{equation}
  S^{t\prime}=\mathrm{Edit}(S^t,u).
\end{equation}
The updated state is saved and becomes the new alignment target. The harness compares it with \textit{Current}, and the coding agent modifies the code using the new differences and implementation bindings, reusing the execution and feedback loop of refinement.

  \section{Experiments}
\label{en:sec:experiments}

We evaluate LiveWorkspace from four perspectives: end-to-end Screenshot-to-Code
performance (\S\ref{en:sec:exp-overall}), the representational capabilities of Shared VS
(\S\ref{en:sec:exp-svs}), the role of visual state in iterative repair
(\S\ref{en:sec:exp-repair}), and controlled user editing
(\S\ref{en:sec:exp-edit}).

\begin{table*}[!t]
\centering
\compacttable

\fitwidth{\textwidth}{
\begin{tabular}{l|ccccc>{\columncolor{black!6}}c|ccccc>{\columncolor{black!6}}c}
\toprule

\textbf{Model}
& \multicolumn{6}{c|}{\textbf{Design2Code}}
& \multicolumn{6}{c}{\textbf{ScreenBench}} \\

\hhline{~------|------}

& \textbf{Block}
& \textbf{Text}
& \textbf{Position}
& \textbf{Color}
& \textbf{CLIP}
& \textbf{Avg}
& \textbf{Block}
& \textbf{Text}
& \textbf{Position}
& \textbf{Color}
& \textbf{CLIP}
& \textbf{Avg} \\

\midrule

\multicolumn{13}{c}{\textbf{Direct Multimodal Models}} \\
\midrule

GPT-4o
& 84.3 & 96.3 & 89.9 & 88.0 & 92.2 & 90.1
& 75.0 & 84.0 & 72.1 & 70.1 & 77.2 & 75.7 \\

GPT-4V
& 83.5 & 95.9 & 89.1 & 86.8 & 90.8 & 89.2
& 71.6 & 81.3 & 69.7 & 68.6 & 75.5 & 73.3 \\

GPT-5.2
& 86.8 & 96.1 & 91.2 & 91.4 & 90.5 & 91.2
& 67.1 & 92.1 & 73.6 & 84.1 & 86.7 & 80.7 \\

Gemini-2.5-Pro
& 88.8 & 93.1 & 85.6 & 71.2 & 81.0 & 83.9
& 71.2 & 91.1 & 73.6 & 82.4 & 88.1 & 81.3 \\

Claude Sonnet 4.6
& \underline{94.5} & 96.1 & 92.2 & 84.4 & 91.5 & 91.7
& 76.7 & \underline{93.0} & 74.2 & 81.5 & 86.6 & 82.4 \\

Claude Opus 4.6
& \textbf{95.1} & 97.3 & 92.5 & 86.1 & 92.2 & 92.6
& \underline{77.2} & 90.7 & \textbf{75.3} & 82.4 & \textbf{90.1} & 83.1 \\

Claude Opus 4.8
& 94.1 & \textbf{98.7} & \underline{92.8} & 88.1 & \underline{92.4} & \underline{93.2}
& 76.9 & \textbf{93.7} & 73.9 & 84.1 & \underline{88.4} & \textbf{83.4} \\

Qwen2.5-VL-72B
& 86.6 & 91.6 & 76.8 & 67.8 & 77.8 & 80.1
& \textbf{77.5} & 78.9 & 61.4 & 51.8 & 63.9 & 66.7 \\

\midrule

\multicolumn{13}{c}{\textbf{Training-free Baselines}} \\
\midrule

DCGen
& 83.7 & 95.7 & 88.9 & 86.0 & 89.9 & 88.8
& 73.5 & 82.6 & 70.8 & 69.5 & 76.2 & 74.5 \\

LaTCoder
& 83.4 & 95.5 & 89.4 & 86.4 & 90.7 & 89.1
& 73.1 & 83.3 & 71.3 & 70.4 & 76.7 & 75.0 \\

LayoutCoder
& 83.6 & 96.8 & 88.9 & 88.0 & 91.1 & 89.7
& 73.7 & 83.8 & 71.4 & 70.4 & 76.6 & 75.2 \\

ScreenCoder (Agentic)
& 86.9 & \underline{97.7} & 92.2 & 90.4 & \textbf{92.6} & 92.0
& \underline{77.2} & 85.8 & \underline{75.0} & 73.7 & 81.1 & 78.6 \\

GPT-5.2 (Matched Budget)
& 87.7 & 96.3 & 91.8 & \underline{92.4} & 90.8 & 91.8
& 67.7 & 92.3 & 74.1 & \underline{85.2} & 86.5 & 81.2 \\

\midrule

\multicolumn{13}{c}{\textbf{Post-Trained Baselines}} \\
\midrule

UI2Code$^{N}$
& 85.0 & 90.5 & 75.8 & 76.5 & 87.3 & 83.0
& 74.6 & 78.9 & 58.1 & 58.1 & 76.0 & 69.1 \\

Websight-8B
& 75.6 & 90.6 & 77.2 & 78.0 & 86.1 & 81.5
& 67.9 & 76.9 & 55.2 & 60.9 & 75.1 & 67.2 \\

ScreenCoder (Finetuned)
& 84.7 & 96.7 & 91.5 & 88.8 & 91.1 & 90.6
& 76.2 & 84.3 & 73.7 & 71.3 & 78.8 & 76.9 \\

\midrule

\multicolumn{13}{c}{\textbf{Ours}} \\
\midrule

\textbf{LiveWorkspace}
& 89.0 & 96.8 & \textbf{93.5} & \textbf{96.2} & 92.3 & \textbf{93.6}
& 70.7 & \textbf{93.7} & 74.5 & \textbf{90.6} & 86.6 & \underline{83.2} \\

\bottomrule
\end{tabular}
}

\par
\normalfont\normalsize
\caption{End-to-end Screenshot-to-Code performance on Design2Code and ScreenBench.}
\label{en:tab:combined_results}
\end{table*}

\subsection{Experimental Setup}
\label{en:sec:exp-setup}

\textbf{Benchmarks and metrics.} We evaluate on the complete test sets of Design2Code and ScreenBench \cite{design2code,screencoder}. We report Block, Text, Position, Color, CLIP, and their mean, Avg \cite{design2code}, using pretrained CLIP image features \cite{clip}. We also use Design2Code for representation and repair analyses. User editing is evaluated separately on 100 synthetic controlled cases. ScreenBench is evaluated after freezing the visual tools, recovery rules, and thresholds to assess cross-benchmark generalization.

\textbf{Baselines.} Baselines include direct multimodal models such as Qwen2.5-VL \cite{qwen25vl}, training-free UI-to-Code methods \cite{dcgen,latcoder,layoutcoder,screencoder}, and models trained for UI-to-Code \cite{ui2coden,websight,screencoder}. Existing methods use their public implementations and recommended configurations.

\textbf{Controlled setting.} The end-to-end LiveWorkspace evaluation uses GPT-5.2 as its sole coding agent. GPT-5.2 Direct is a single-call reference. All methods use the same browser environment, and invalid or unrenderable outputs remain in the evaluation.

\textbf{End-to-end protocol.} LiveWorkspace constructs Recovered \textit{Target} Shared VS, while GPT-5.2 (Matched Budget) uses free-text visual analysis; both then generate an initial implementation. Both use two initialization calls and up to three repair rounds, with Shared VS feedback for LiveWorkspace and screenshot feedback for the control. We match actual model-call counts and stagewise budget limits, including browser-execution budgets. Both allow early termination and are evaluated on their last generated implementation.

Appendix~\ref{en:app:budget-pilot} reports initial-generation model calls, token usage, and visual-tool latency.

\subsection{Overall End-to-End Performance}
\label{en:sec:exp-overall}

Table~\ref{en:tab:combined_results} reports final reconstruction performance of the complete LiveWorkspace harness under the end-to-end protocol in \S\ref{en:sec:exp-setup}.

With matched model-call counts and budget limits, LiveWorkspace improves Avg from 91.8 to 93.6 on Design2Code and from 81.2 to 83.2 on ScreenBench, gains of 1.8 and 2.0 points. These gains support the shared-state approach under matched call counts and budget limits. On ScreenBench, the largest gains over the matched control are in Color (85.2 to 90.6) and Block (67.7 to 70.7).

LiveWorkspace also achieves competitive reconstruction scores relative to the direct multimodal models and existing UI-to-Code systems in Table~\ref{en:tab:combined_results}.

\begin{samepage}
\textbf{Two-call generation control.} On a 50-page Design2Code subset, we compare free-text analysis with \textit{Recovered} \textit{Target} Shared VS for initial generation. Both use two GPT-5.2 calls with matched stagewise token limits and no refinement. Shared VS achieves 94.53 Avg, compared with 91.13 for free-text analysis, a gain of 3.40 points. Appendix~\ref{en:app:controlled-experiments} details this control.
\par
\end{samepage}

\subsection{Shared VS: Sufficiency, Recoverability, and Transferability}
\label{en:sec:exp-svs}

We compare two construction sources for \textit{Target}: \textit{Recovered} \textit{Target} Shared VS is constructed from the target screenshot, whereas \textit{Reference} \textit{Target} Shared VS is derived from the reference implementation.

\begin{figure}[!t]
\centering
\includegraphics[width=\columnwidth]{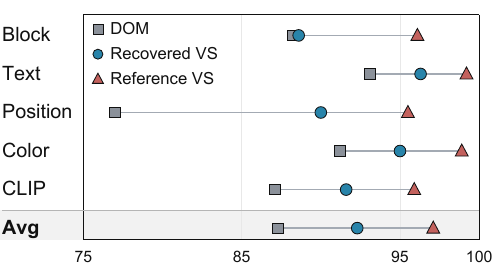}
\caption{Screenshot-free initial generation on Design2Code from different visual representations using GPT-5.2.}
\label{en:fig:representation-sufficiency}
\end{figure}

\textbf{Representation sufficiency.} We fix GPT-5.2 as the coder and compare initial generation from different representations without giving the coder the target screenshot. \textit{Reference} \textit{Target} Shared VS is an oracle representation deterministically derived from the reference rendered page into the same canonical schema; the coder receives only the resulting state. In Figure~\ref{en:fig:representation-sufficiency}, screenshot-inferred DOM, \textit{Recovered} Shared VS, and \textit{Reference} Shared VS achieve 87.3, 92.3, and 97.1 Avg, respectively. \textit{Recovered} Shared VS improves over DOM by 5.0 points, supporting its ability to retain visual information for code generation.

\textbf{Recoverability.} Table~\ref{en:tab:recovery} shows strong text and color recovery, while hierarchy attachment, size estimation, and layout-primitive recognition remain the main bottlenecks. Visual grouping and layout direction are recovered more consistently.

Agreement with a reference decomposition does not fully measure the completeness of recovered visual constraints: the same arrangement may be described as a grid or as a vertical sequence of horizontal sequences. Figure~\ref{en:fig:representation-sufficiency} therefore provides complementary downstream evidence. Its 4.8-point \textit{Recovered}--\textit{Reference} gap captures the aggregate effect of state recovery on reconstruction, rather than layout-primitive mismatch alone.

\begin{table}[!t]
\centering
\compacttable
\begin{tabularx}{\columnwidth}{*{3}{>{\centering\arraybackslash}X}}
\toprule
Component & Metric & Result \\
\midrule
\rowcolor{black!6}
 & Co-group F1 $\uparrow$ & 0.79 \\
\rowcolor{black!6}
\multirow{-2}{*}{Hierarchy} & Hierarchy Acc. $\uparrow$ & 50.6 \\
Geometry & Size within tol. $\uparrow$ & 41.2 \\
\rowcolor{black!6}
 & Primitive F1 $\uparrow$ & 0.44 \\
\rowcolor{black!6}
\multirow{-2}{*}{Layout} & Direction Acc. $\uparrow$ & 89.8 \\
 & Coverage $\uparrow$ & 96.8 \\
\multirow{-2}{*}{Text} & Exact Match $\uparrow$ & 93.3 \\
\rowcolor{black!6}
 & Coverage $\uparrow$ & 98.2 \\
\rowcolor{black!6}
\multirow{-2}{*}{Color} & Exact Match $\uparrow$ & 86.7 \\
\bottomrule
\end{tabularx}
\par
\normalfont\normalsize
\caption{Screenshot-to-\textit{Target} Shared VS recovery by component. F1 scores are on a 0--1 scale; other results are percentages.}
\label{en:tab:recovery}
\end{table}

\textbf{Cross-model transferability.} We construct \textit{Recovered} \textit{Target} Shared VS using GPT-5.2 and the harness, then reuse the state across coders for initial generation without refinement. As shown in Table~\ref{en:tab:transfer}, Shared VS improves DeepSeek-V4's Avg from 89.6 to 92.7 on Design2Code and from 79.7 to 83.0 on ScreenBench compared with Raw Visual Observations. These results support reuse of the constructed state across coding models.

\begin{table}[H]
\centering
\compacttable
\begingroup
\fitwidth{\columnwidth}{\begin{tabular}{l>{\raggedright\arraybackslash}p{0.40\columnwidth}ccc}
\toprule
Coder & Input & Img & D2C & SB \\
\midrule
GPT-5.2 & Screenshot & \ding{51} & 91.2 & 80.7 \\
GPT-5.2 & Recovered Shared VS & \ding{55} & 92.3 & 83.1 \\
DeepSeek-V4 & Raw Visual $O$ & \ding{55} & 89.6 & 79.7 \\
DeepSeek-V4 & Recovered Shared VS & \ding{55} & 92.7 & 83.0 \\
\bottomrule
\end{tabular}}
\endgroup
\par
\normalfont\normalsize
\caption{Cross-model transfer of \textit{Recovered} \textit{Target} Shared VS for initial generation. Shared VS and Raw Visual \(O\) are text-only inputs.}
\label{en:tab:transfer}
\end{table}

\textbf{Component ablation.} Table~\ref{en:tab:hal} removes hierarchy, attributes, or layout from \textit{Reference} \textit{Target} Shared VS. Removing attributes produces the largest drop, from 97.1 to 62.5 Avg; removing hierarchy or layout reduces Avg to 83.1 or 91.2, respectively.

\begin{table}[H]
\centering
\compacttable
\begin{tabularx}{\columnwidth}{*{3}{>{\centering\arraybackslash}X}}
\toprule
Representation & Components & Avg $\uparrow$ \\
\midrule
$-$ Hierarchy  & $A+L$ & 83.1 \\
$-$ Attributes & $H+L$ & 62.5 \\
$-$ Layout     & $H+A$ & 91.2 \\
\textbf{Full} & $H+A+L$ & \textbf{97.1} \\
\bottomrule
\end{tabularx}
\par
\normalfont\normalsize
\caption{Initial-generation ablation of \textit{Reference} \textit{Target} Shared VS on Design2Code with GPT-5.2. Full denotes \(H+A+L\).}
\label{en:tab:hal}
\end{table}

\subsection{Iterative Repair in LiveWorkspace}
\label{en:sec:exp-repair}

\textbf{Structured feedback.} Table~\ref{en:tab:repair} compares feedback interfaces for a single round of repair from the same \(C_0\) using GPT-5.2. Repair Rate measures the proportion of initial Shared VS discrepancies resolved, while Final Fidelity uses the independent Design2Code evaluator described in Appendix~\ref{en:app:repair-metrics}. Adding Structured Diff to \textit{Target} and \textit{Current} feedback raises Final Fidelity from 90.0 to 91.9 and Repair Rate from 11.8\% to 46.3\%, turning state differences into explicit repair targets.

\textbf{Implementation binding.} Adding source binding raises Final Fidelity from 91.9 to 93.1 and Repair Rate from 46.3\% to 53.8\%. Diff identifies visual discrepancies, while binding provides implementation-location cues. Collateral rises from 4.9 to 5.1 newly defective nodes per page, but remains below Screenshot Feedback (14.6) and Raw Observations (13.7).

\begin{table}[H]
\centering
\begingroup
\definecolor{repairdiff}{HTML}{397DA8}
\renewcommand{\fidelityscore}[2]{#1\rlap{\textsubscript{\textcolor{repairdiff}{\normalfont\footnotesize\makebox[1.7em][r]{#2}}}}}
\small
\setlength{\tabcolsep}{1.5pt}
\renewcommand{\arraystretch}{1.08}
\begin{tabularx}{\columnwidth}{@{}>{\raggedright\arraybackslash}X *{4}{@{\hspace{2pt}}c} @{\hspace{2pt}}c @{\hspace{2pt}}c @{\hspace{10pt}}c@{}}
\toprule
\multirow{2}{*}{Condition} & \multicolumn{4}{c}{Feedback} & \multicolumn{3}{c}{Results} \\
\cmidrule(lr){2-5}\cmidrule(l){6-8}
& $T$ & $C$ & $\Delta$ & $B$
& Repair $\uparrow$
& Fidelity $\uparrow$
& Collateral $\downarrow$ \\
\midrule
Initial $C_0$ & & & & & -- & \fidelityscore{91.2}{0.0} & -- \\
Screenshot & & & & & 34.0 & \fidelityscore{91.6}{+0.4} & 14.6 \\
Raw $O$ & & & & & 41.2 & \fidelityscore{89.8}{-1.4} & 13.7 \\
Static VS & \ding{51} & & & & \phantom{0}9.9 & \fidelityscore{90.9}{-0.3} & \phantom{0}\textbf{1.3} \\
Target + Current & \ding{51} & \ding{51} & & & 11.8 & \fidelityscore{90.0}{-1.2} & \phantom{0}2.4 \\
+ Diff & \ding{51} & \ding{51} & \ding{51} & & 46.3 & \fidelityscore{91.9}{+0.7} & \phantom{0}4.9 \\
\rowcolor{black!5}
\textbf{LiveWorkspace} & \ding{51} & \ding{51} & \ding{51} & \ding{51} & \textbf{53.8} & \fidelityscore{\textbf{93.1}}{+1.9} & \phantom{0}5.1 \\
\bottomrule
\end{tabularx}
\endgroup
\par
\normalfont\normalsize
\caption{Single round repair with GPT-5.2 from a shared \(C_0\). $T$, $C$, $\Delta$, and $B$ denote \textit{Target}, \textit{Current}, Structured Diff, and binding; $O$ denotes Raw Observations. Repair Rate (\%) measures resolved initial Shared VS discrepancies; Final Fidelity is Design2Code Avg; Collateral counts newly defective \textit{Target} nodes per page. Subscripts indicate fidelity changes from \(C_0\).}
\label{en:tab:repair}
\end{table}

On 50 pages containing 1,392 visual nodes, binding coverage is 100\%, with 78.9\% independent position agreement. The manual audit in Appendix~\ref{en:app:binding-audit} covers 250 nodes and finds 73.2\% actionable source localization.

\begin{figure*}[!t]
\centering
\includegraphics[width=\textwidth]{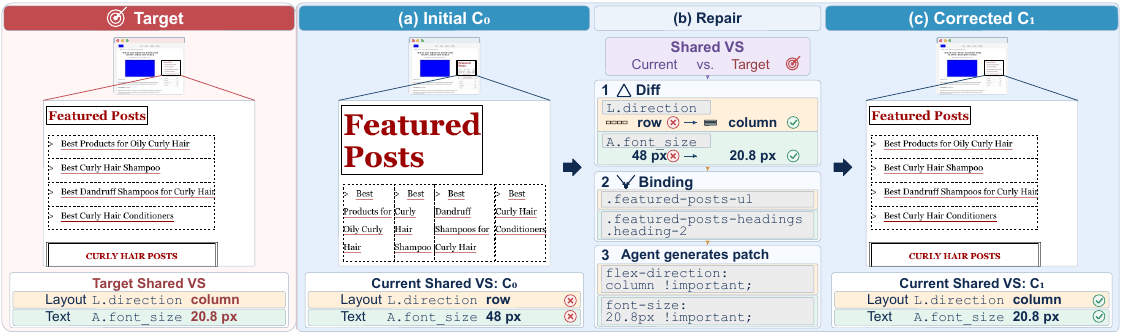}
\caption{Automated repair with LiveWorkspace. \textit{Target}--\textit{Current} Shared VS differences expose layout-direction and font-size errors. Implementation bindings connect these errors to code locations, enabling the agent to generate a patch that restores the target layout and typography.}
\label{en:fig:qualitative-repair}
\end{figure*}

Figure~\ref{en:fig:qualitative-repair} illustrates how Shared VS differences and implementation bindings guide an agent-generated patch to correct layout and typography errors.

\begin{figure}[!t]
\centering
\includegraphics[width=\columnwidth]{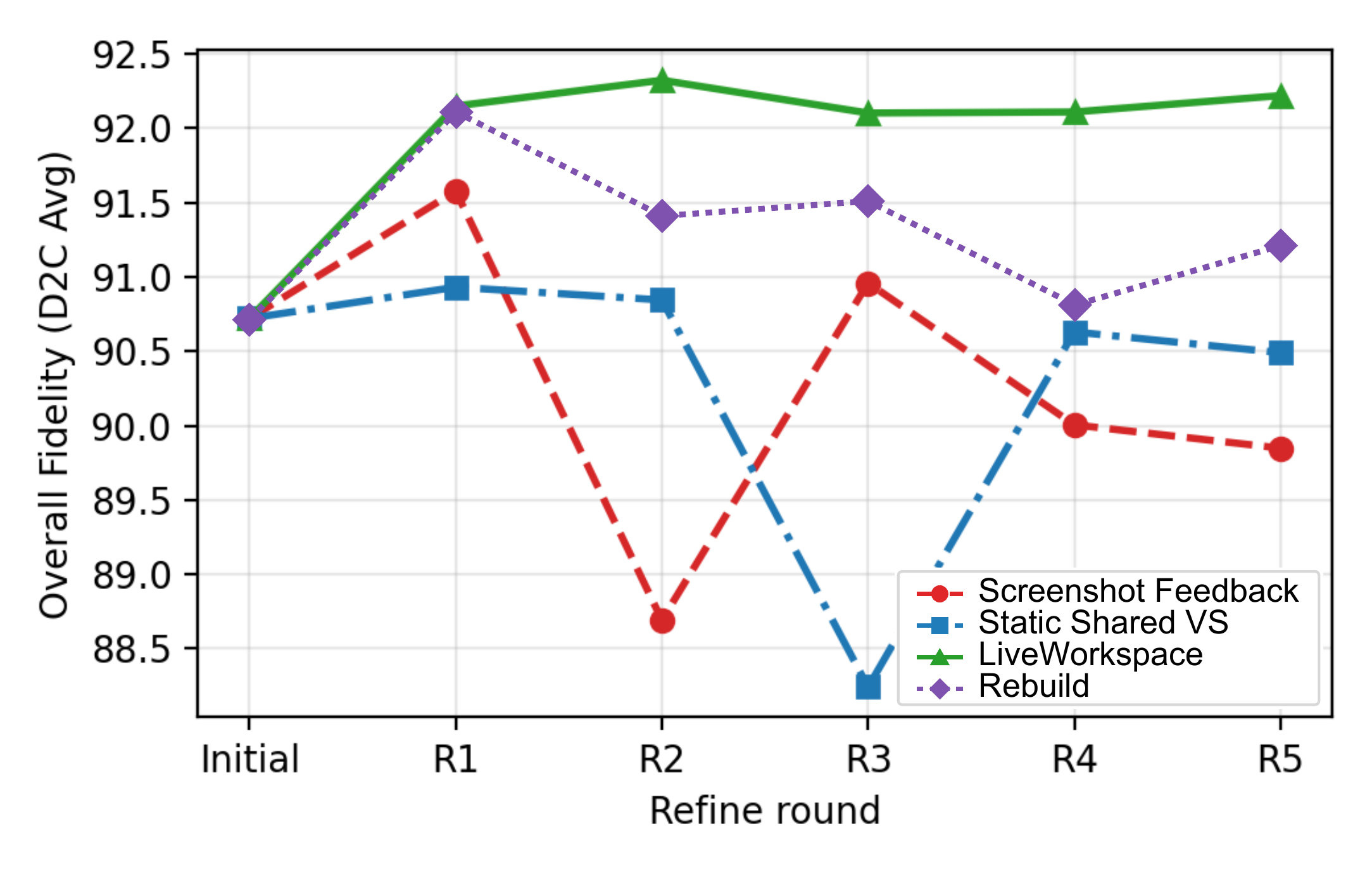}
\caption{Five-round repair on a shared 50-page Design2Code subset with GPT-5.2. All conditions start from the same \(C_0\); LiveWorkspace retains \textit{Target} Shared VS, while Rebuild reconstructs it before each round.}
\label{en:fig:repair-curve}
\label{en:fig:target-lifecycle}
\end{figure}

\textbf{Target persistence.} Figure~\ref{en:fig:repair-curve} compares five repair rounds on a 50-page Design2Code subset using GPT-5.2, with all four conditions sharing an initial \(C_0\) of 90.7 Avg. Rebuild uses the same LiveWorkspace feedback interface and initial \textit{Target}, but reconstructs \textit{Target} before each round under matched repair budgets and token limits. LiveWorkspace maintains fidelity around 92.2 after the first-round improvement, while Rebuild finishes at 91.2; Screenshot Feedback and Static Shared VS fluctuate in later rounds. Across the five rounds, retaining \textit{Target} uses 42.7\% fewer input tokens and 15.7\% fewer output tokens in mean per-page totals, as reported in Appendix~\ref{en:app:controlled-experiments}.

\subsection{Controlled Editing in LiveWorkspace}
\label{en:sec:exp-edit}

LiveWorkspace updates and saves \textit{Target} Shared VS according to the user instruction, then reuses the alignment-and-repair loop from \S\ref{en:sec:exp-repair}. The harness recovers \textit{Current} after execution to verify the edit.

Visual GUI agents study instruction-to-element grounding \cite{seeclick,uground,osatlas,showui}. Our controlled cases test this grounding for localized code edits.

We construct 100 controlled cases covering Appearance, Layout, and Structure. Each page contains three or four repeated instances whose source and visual orders differ, and instructions identify the target by visual position. Both methods receive the same inputs and use matched budget limits.

Edit Success requires completing the requested change on the intended instance without modifying other repeated instances. Instance Preservation measures unchanged non-target repeated instances; Global Preservation measures unchanged non-target node presence, text, color, and geometry/layout state. Both preservation metrics are computed per case and then averaged. All reordering cases undergo independent rendered-page verification of the final visual order, as detailed in Appendix~\ref{en:app:editing-protocol}.

In Table~\ref{en:tab:edit}, LiveWorkspace improves Edit Success from 41.0\% to 93.0\%, Instance Preservation from 76.5\% to 97.5\%, and Global Preservation from 93.9\% to 99.2\%. These results support reusing the harness's state-alignment interface for localized editing while preserving non-target content and layout.

\begin{table}[H]
\centering

\compacttable
\fitwidth{\columnwidth}{\begin{tabular}{lccc}
\toprule
Method & Edit Success $\uparrow$ & Inst.\ Pres.\ $\uparrow$ & Global Pres. $\uparrow$ \\
\midrule
Direct Editing        & 41.0 & 76.5 & 93.9 \\
\textbf{LiveWorkspace} & \textbf{93.0} & \textbf{97.5} & \textbf{99.2} \\
\bottomrule
\end{tabular}}
\par
\normalfont\normalsize
\caption{Controlled editing across 100 Appearance, Layout, and Structure cases. All results are percentages. Global Preservation covers non-target node presence, text, color, and geometry/layout.}
\label{en:tab:edit}
\end{table}

  \section{Conclusion}
\label{en:sec:conclusion}

We introduce LiveWorkspace, a Screenshot-to-Code harness that externalizes visual understanding as Shared Visual State (Shared VS). LiveWorkspace retains \textit{Target} and recovers \textit{Current} after execution in a common, implementation-independent schema, with Structured Diff and implementation bindings guiding code modifications. With GPT-5.2 and matched model-call counts and budget limits, LiveWorkspace improves end-to-end reconstruction. Further experiments demonstrate cross-model state reuse and improvements in iterative repair and controlled editing. Controlled comparisons highlight complementary benefits of structuring visual evidence under matched budgets and retaining target intent instead of reconstructing it across iterations. These findings support treating visual understanding as a persistent shared working state throughout code generation, execution, and modification.

  \section*{Limitations}

LiveWorkspace has several limitations. First, recovering Target Shared VS from
screenshots remains imperfect, particularly in visual hierarchy, geometric estimation,
and layout relations. These errors can propagate to subsequent generation and repair.
Second, Target--Current alignment relies on normalized text matching and greedy spatial
correspondence. While efficient, these rules may produce incorrect matches when text
changes, elements move substantially, or page structures differ considerably.

Our evaluation focuses on visual reconstruction of webpages and controlled user edits
involving appearance, layout, and structure. It does not directly assess interactive behavior or multi-page/full-website functionality \cite{interaction2code,mrweb,webgenbench}, responsive adaptation, or longer-term development workflows. In addition,
recovering Current Shared VS relies on browser runtime information, including DOM,
computed styles, and rendered geometry. Extending this approach to interfaces without
comparable runtime access would require alternative state-recovery mechanisms.

  \bibliography{references}
  \clearpage
  \appendix
  \renewcommand{\theHsection}{enapp.\Alph{section}}
  \renewcommand{\theHsubsection}{enapp.\Alph{section}.\arabic{subsection}}
  \section{State and Agent Interfaces}
\label{en:app:state-interface}

Target and Current denote runtime roles; Recovered and Reference distinguish Target
construction sources. All instances use the same \(S=(H,A,L)\) schema.

\subsection{Visual State and Recovery Evidence}

\textbf{Visual state.} \(H\) stores grouping and containment; \(A\) stores observable
geometry, text, and paint; \(L\) stores within-group arrangements, direction, spacing,
and padding. Synthetic groups need not correspond to single source elements, so the
visual hierarchy need not be isomorphic to the DOM.

\textbf{Target recovery evidence.} Recovered Target is constructed entirely from the screenshot, screenshot-derived region, text, and paint observations \(O\), and the inferred hierarchy. GPT-5.2 infers hierarchy; the harness recovers attributes and layout from this evidence, without accessing reference DOM, computed styles, or reference-code rendering. Reference Target is deterministically recovered from the reference rendered page or implementation, using runtime DOM, computed styles, and painted geometry as evidence. The reconstruction coder receives only the resulting state.

\textbf{Current recovery evidence.} Current uses runtime DOM, computed styles, and
painted geometry. Canonicalization collapses non-visual wrappers, prunes invisible
elements, and infers layout from painted geometry. Tags, classes, selectors, and source
locations remain implementation evidence or independent binding, outside Shared VS.

\textbf{Implementation binding.} For self-generated, single-file HTML, the binder parses source elements and greedily assigns visual nodes one-to-one in document order, using tag compatibility and containment of the node text in the candidate element's full \texttt{get\_text()} output. Binding metadata remains separate from Shared VS. The position audit in Appendix~\ref{en:app:binding-audit} evaluates these assignments independently.

\subsection{Correspondence and State Lifecycle}

Node IDs are independent. Text-bearing nodes match by normalized exact text, with
center distance resolving ties. Remaining nodes match greedily and one-to-one by
absolute bounding-box IoU (\(IoU\geq0.5\)), prioritizing larger overlaps. Attributes are
compared only for matched pairs; unmatched reportable Target and Current nodes are
\textbf{MISSING} and \textbf{EXTRA}, respectively.

Target is retained after construction. Each execution recovers \(S^c_k\), recomputes
\(\Delta_k\), and refreshes \(B_k\). Refinement reuses Target; editing updates Target
before reusing alignment. Persistence concerns the external target state, while
Current is recovered from each execution.

\subsection{Agent Inputs and Experimental Conditions}

\textbf{Initial generation.} Construction uses the screenshot and \(O\); code generation
uses Target. Appendix~\ref{en:app:budget-pilot} checks screenshot and reference-HTML
access in the pilot \(C_0\) requests.

\textbf{Full repair interface.} The agent receives \(C_k,S^t,S^c_k,\Delta_k,B_k\)
and decides on local patches or broader changes. The harness provides recovered states
and execution feedback. Table~\ref{en:tab:repair} progressively adds states, Diff, and
binding to compare interfaces.

\textbf{User editing.} The agent locates affected Target nodes and attributes, updates and saves \(S^{t'}\), and reuses the full repair interface. Each execution recovers Current and refreshes Diff and binding against the updated Target.

\textbf{Representation conditions.} Figure~\ref{en:fig:representation-sufficiency} and Table~\ref{en:tab:transfer} evaluate initial-generation outputs without subsequent repair. They use the same benchmark pages as Table~\ref{en:tab:combined_results}, which instead reports the final outputs of the complete LiveWorkspace runs. Their results are therefore reported separately by evaluation stage. Full in Table~\ref{en:tab:hal} uses Reference Target Shared VS and corresponds to the Reference condition in Figure~\ref{en:fig:representation-sufficiency}, rather than the Recovered condition in the main table.

\subsection{Implementation Details}

\textbf{Visual observations.} Region candidates are obtained with PP-DocLayoutV3 \cite{paddleocrvl15}. Text detection and recognition use PP-OCRv6\_medium\_det and PP-OCRv6\_medium\_rec \cite{ppocrv6}, respectively, from the PaddleOCR toolkit \cite{paddleocr3}. Paint observations are extracted by \texttt{h2\_m1c\_paint.py}, which uses screenshot color connected components to recover painted non-text geometry, colors, and borders. These local visual tools run on CPU. GPT-5.2 separately infers hierarchy from the screenshot and the resulting observations.

\textbf{Canonicalization.} Non-painting wrappers with no text of their own and exactly one child are collapsed; wrapper candidates include \texttt{div}, \texttt{span}, \texttt{section}, \texttt{article}, \texttt{main}, \texttt{header}, and \texttt{footer}. Elements such as \texttt{picture} are treated as non-painting wrappers. Invisible and off-canvas elements are pruned, with a 0.5 px collapsed-element threshold. Sequence, grid, overlay, and freeform layout relations, together with direction, gap, and padding, are inferred from child geometry; layout primitives are not assigned from CSS \texttt{display} labels.

\textbf{Diff tolerances.} Width and height discrepancies are reported beyond \(\max(2\,\mathrm{px},0.05\times\text{target size})\). Position discrepancies are reported for center differences greater than 12 px, and gap uses a 6 px tolerance. Colors are parsed into integer RGB triples and compared exactly. Font family is not included in Diff.

\subsection{Worked State and Diff Examples}

The cards below reproduce a root-container node, a text-leaf node, and a Structured Diff excerpt. The Diff compares \(C_0\) with Recovered Target Shared VS for Design2Code case 341. It reports 29 Target nodes, 30 Current nodes, 29 aligned pairs, and property differences on 22 nodes; only the displayed portion is reproduced. \texttt{None} denotes an absent value in this output.

\sharedexamplecard{Root-container node}{\sharedexampleline{0.0pt}{\{}
\sharedexampleline{4.4pt}{"id": "G",}
\sharedexampleline{4.4pt}{"type": "region",}
\sharedexampleline{4.4pt}{"role": "visual\_container",}
\sharedexampleline{4.4pt}{"attributes": \{}
\sharedexampleline{8.8pt}{"geometry": \{"width": 1280.0, "height": 720.0\},}
\sharedexampleline{8.8pt}{"appearance": \{"fill": "rgb(51, 63, 72)"\},}
\sharedexampleline{8.8pt}{"typography": \{"color": "rgb(255, 255, 255)"\}}
\sharedexampleline{4.4pt}{\},}
\sharedexampleline{4.4pt}{"layout": \{}
\sharedexampleline{8.8pt}{"type": "sequence",}
\sharedexampleline{8.8pt}{"direction": "column",}
\sharedexampleline{8.8pt}{"padding": [0.0, null, null, null],}
\sharedexampleline{8.8pt}{"gap": 17.0,}
\sharedexampleline{8.8pt}{"wrap": false}
\sharedexampleline{4.4pt}{\}}
\sharedexampleline{0.0pt}{\}}
}
\sharedexamplecard{Text-leaf node}{\sharedexampleline{0.0pt}{\{}
\sharedexampleline{4.4pt}{"id": "t0",}
\sharedexampleline{4.4pt}{"type": "element",}
\sharedexampleline{4.4pt}{"role": "text",}
\sharedexampleline{4.4pt}{"attributes": \{}
\sharedexampleline{8.8pt}{"geometry": \{"width": 334.0, "height": 18.0\},}
\sharedexampleline{8.8pt}{"typography": \{"color": "rgb(255, 255, 255)"\},}
\sharedexampleline{8.8pt}{"content": \{"text": "SOUTH ASIA INSTITUTE | College of Liberal Arts"\}}
\sharedexampleline{4.4pt}{\},}
\sharedexampleline{4.4pt}{"children": []}
\sharedexampleline{0.0pt}{\}}
}
\sharedexamplecard{Structured Diff · case 341 (excerpt)}{\sharedexampleline{0.0pt}{\# Visual-State Diff (target 29 nodes, current 30, aligned 29)}
\vspace{3pt}
\sharedexampleline{0.0pt}{\#\# PROPERTY differences (22 nodes)}
\sharedexampleline{4.4pt}{node visual\_container at (0.0, 0.0, 1280.0, 720.0):}
\sharedexampleline{13.200000000000001pt}{typography.color: target='rgb(255, 255, 255)'  current=None}
\sharedexampleline{4.4pt}{node visual\_container at (0.0, 0.0, 1280.0, 36.0):}
\sharedexampleline{13.200000000000001pt}{typography.color: target='rgb(255, 255, 255)'  current=None}
\sharedexampleline{4.4pt}{node visual\_container at (51.0, 213.0, 1106.0, 306.0):}
\sharedexampleline{13.200000000000001pt}{position.x: target=51.0  current=87}
\sharedexampleline{13.200000000000001pt}{typography.color: target='rgb(255, 255, 255)'  current=None}
\sharedexampleline{4.4pt}{node visual\_container at (51.0, 213.0, 240.0, 135.0):}
\sharedexampleline{13.200000000000001pt}{position.x: target=51.0  current=87}
}

\section{Initial-Generation Budget Pilot}
\label{en:app:budget-pilot}

We measured initial-generation budgets on 50 Design2Code pages. Direct makes one GPT-5.2 call per page. LiveWorkspace makes two calls: Target hierarchy inference and generation of \(C_0\) from the resulting Target Shared VS. All stages use GPT-5.2. Table~\ref{en:tab:budget-pilot} illustrates pages 341, 1002, and 10032 from this study, reporting actual input/output usage returned by the API, configured token caps, and call latency.

This pilot measures Target construction and initial code generation only; it does not include the automatic repair rounds used in the main-table evaluation.

\begin{table}[!htbp]
\centering
\compacttable
\fitwidth{\columnwidth}{\begin{tabular}{llrrrr}
\toprule
Case & Stage & Input & Output & Cap & Time \\
\midrule
341 & Direct & 1,263 & 1,127 & 32,000 & 25.2 \\
 & Target & 2,219 & 349 & -- & 9.9 \\
 & $C_0$ & 5,518 & 2,457 & 96,000 & 21.5 \\
\midrule
1002 & Direct & 1,263 & 856 & 32,000 & 23.6 \\
 & Target & 1,944 & 420 & -- & 10.7 \\
 & $C_0$ & 3,727 & 1,658 & 96,000 & 15.2 \\
\midrule
10032 & Direct & 1,263 & 1,731 & 32,000 & 49.8 \\
 & Target & 2,286 & 527 & -- & 13.1 \\
 & $C_0$ & 6,521 & 2,846 & 96,000 & 22.2 \\
\bottomrule
\end{tabular}}
\par
\normalfont\normalsize
\caption{API usage and model-call latency for three illustrative pages from the 50-page initial-generation budget study. All calls use GPT-5.2.}
\label{en:tab:budget-pilot}
\end{table}

\textbf{Accounting.} Each row corresponds to one model call; Time is in seconds. Target
denotes hierarchy inference, with subsequent attribute and layout recovery performed by
the harness. Direct and \(C_0\) use token caps of 32,000 and 96,000, respectively. The
Target-stage cap was not recorded in the pilot summary and is shown as a dash. The logs
recorded no retries and no rendering of initial code \(C_0\) on these three pages.

\textbf{Per-page means.} Across the 50 pages, Direct uses one call per page, averaging approximately 1.26K input tokens and 1.24K output tokens. LiveWorkspace's two stages together use two calls, averaging approximately 7.51K input tokens and 2.81K output tokens. Target hierarchy inference alone averages approximately 2.19K input and 0.45K output tokens; \(C_0\) generation averages approximately 5.32K input and 2.36K output tokens. This study compares the initial-generation costs of GPT-5.2 Direct and LiveWorkspace, which use the same model but different call counts and token budgets. The main-table GPT-5.2 (Matched Budget) control instead matches actual model-call counts and stagewise budget limits.

\textbf{Context checks.} The metadata for all three \(C_0\) requests records no image
attachment and no mention of reference HTML. These checks concern the separate
code-generation call; hierarchy construction still uses the target screenshot.

\textbf{Model and tool latency.} Across the 50 pages, Direct's mean model-call latency is 33.1 seconds. Target hierarchy inference and \(C_0\) generation average 11.6 and 20.3 seconds, respectively; their sum is 31.9 seconds for LiveWorkspace. These records use cached low-level visual observations. Separately measured M1 processing averages 10.2 seconds. Adding it gives a serial initial-generation estimate of \(31.9+10.2=42.1\) seconds. This is a sum of stage means, rather than a directly timed end-to-end latency.

\section{Implementation-Binding Audit}
\label{en:app:binding-audit}

\subsection{Coverage, Consistency, and Position Agreement}

The binding audit covers 50 generated pages with 1,392 rendered visual nodes, comprising 761 text-bearing and 631 non-text nodes. As summarized in Table~\ref{en:tab:binding-audit}, we distinguish binding availability, text/tag consistency, independent position agreement, and manually assessed actionable source localization.

\textbf{Availability and consistency.} All 1,392 nodes receive bindings. For all 761 text-bearing nodes, the bound element's full text contains the corresponding normalized node text; all 631 non-text nodes pass tag-consistency checks. These checks establish availability and text/tag consistency but do not independently identify the intended rendered element.

\textbf{Independent position audit.} Each bound element receives a unique \texttt{data-bchk} marker before re-rendering. Its \texttt{getBoundingClientRect()} box is compared with the corresponding Current Shared VS node's painted box. Both use document-absolute coordinates at a viewport width of 1280 px and zero scroll offset. Position agreement requires the L1 distance between box centers to be at most 24 px. This signal is independent of the binder's tag, text, and document-order checks. Agreement is 1,098/1,392 (78.9\%), comprising 678/761 text-bearing nodes (89.1\%) and 420/631 non-text nodes (66.6\%). This criterion measures spatial agreement rather than exact element identity or source-span correctness.

\begin{table}[!htbp]
\centering
\compacttable
\begin{tabularx}{\columnwidth}{>{\raggedright\arraybackslash}Xrr}
\toprule
Check & Count & Rate (\%) \\
\midrule
Binding coverage & 1392/1392 & 100.0 \\
Text consistency & 761/761 & 100.0 \\
Tag consistency (non-text) & 631/631 & 100.0 \\
Position agreement & 1098/1392 & 78.9 \\
\quad Text nodes & 678/761 & 89.1 \\
\quad Non-text nodes & 420/631 & 66.6 \\
Source-line availability & 1392/1392 & 100.0 \\
\midrule
Actionable source localization & 183/250 & 73.2 \\
\quad Text nodes & 114/135 & 84.4 \\
\quad Non-text nodes & 69/115 & 60.0 \\
\bottomrule
\end{tabularx}
\par
\normalfont\normalsize
\caption{Implementation-binding audit on 50 pages. Position agreement is evaluated independently of text/tag consistency; actionable source localization is assessed by manual review.}
\label{en:tab:binding-audit}
\end{table}

\subsection{Failure Modes and Source Locations}

In the three-page illustrative audit, mismatches mainly involve containers and text--container correspondences. Wrapper
collapse or text promotion during canonicalization can shift visual-node order relative
to source-element order, binding a node to a neighboring element or its parent container.
Of 18 flagged bindings, 11 have center distances exceeding 100 px; smaller deviations
also include tolerance-boundary cases around 28 px.

One audited text node has painted box \((0,545,195,19)\), while its bound element has
box \((0,545,195,76)\), expressed as \((x,y,w,h)\) in pixels. The boxes share an origin,
but the latter covers a taller container region. Text containment and nearby centers can
therefore still select a parent container rather than the intended text element.

All 1,392 bound elements have source-line metadata. A separate manual audit of 250 nodes reports actionable source localization of 183/250 (73.2\%), comprising 114/135 text nodes (84.4\%) and 69/115 non-text nodes (60.0\%). A binding is actionable if it identifies a specific source element or style rule where the intended visual change can be made without further global source search. Source-line availability, position agreement, and actionable source localization are reported separately.

Table~\ref{en:tab:repair} measures repair gains from adding binding, while this audit
characterizes location cues in the 50-page audit. Together they support using
binding as implementation-location cues for the agent. The full repair interface also
retains visual nodes, Current state, and Structured Diff, rather than treating binding as
a guarantee of exact localization.

\section{Additional Controlled Experiments}
\label{en:app:controlled-experiments}

We report controlled Design2Code experiments, using GPT-5.2 and Design2Code Avg to measure reconstruction quality. The free-text versus Shared VS generation control uses 50 Design2Code pages. The four conditions in Figure~\ref{en:fig:repair-curve} share a 50-page Design2Code subset and its initial implementations.

\subsection{Free-Text Analysis versus Shared VS}

We compare free-text analysis with Recovered Target Shared VS as intermediate
interfaces. Both conditions use two model calls per page: intermediate-representation
construction and code generation. Stagewise token limits and budgets are matched,
and actual usage remains below the limits.

In Table~\ref{en:tab:analysis-control}, Recovered Target Shared VS achieves 94.53 Avg
versus 91.13 for free-text analysis, a \textbf{3.40}-point advantage. This comparison
supports the value of Shared VS as a generation interface with matched call counts
and token limits.

\begin{table}[H]
\centering
\compacttable
\setlength{\tabcolsep}{4pt}
\renewcommand{\arraystretch}{1.12}
\begin{tabularx}{\columnwidth}{>{\raggedright\arraybackslash}X >{\raggedleft\arraybackslash}p{0.19\columnwidth} >{\raggedleft\arraybackslash}p{0.19\columnwidth}}
\toprule
\textbf{Intermediate input} & \textbf{Avg} $\uparrow$ & \textbf{Gain} \\
\midrule
Free-text analysis & 91.13 & -- \\
\rowcolor{black!5}
\textbf{Recovered Shared VS} & \textbf{94.53} & \textbf{+3.40} \\
\bottomrule
\end{tabularx}
\par
\normalfont\normalsize
\caption{Initial generation from free-text analysis or Recovered Target Shared VS, without refinement. Both conditions use two GPT-5.2 calls per page with matched stagewise token limits.}
\label{en:tab:analysis-control}
\end{table}

\subsection{Retained versus Reconstructed Target}

We compare two Target lifecycles: LiveWorkspace retains the initial Target, whereas Rebuild reconstructs Target before each repair round. Both conditions share the same \(C_0\) and initial Target, with matched repair budgets and token limits; actual usage remains below the limits. Figure~\ref{en:fig:repair-curve} reports the shared initial Avg and results after repair rounds 1--5; Table~\ref{en:tab:target-token-usage} reports the recorded mean per-page total token usage across the five rounds.

\begin{table}[H]
\centering
\compacttable
\setlength{\tabcolsep}{4pt}
\renewcommand{\arraystretch}{1.12}
\begin{tabularx}{\columnwidth}{>{\raggedright\arraybackslash}X >{\raggedleft\arraybackslash}p{0.19\columnwidth} >{\raggedleft\arraybackslash}p{0.19\columnwidth}}
\toprule
\textbf{Target lifecycle} & \textbf{Input} $\downarrow$ & \textbf{Output} $\downarrow$ \\
\midrule
Rebuild & 27,505 & 15,472 \\
\rowcolor{black!5}
\textbf{Persist} & \textbf{15,774} & \textbf{13,048} \\
\bottomrule
\end{tabularx}
\par
\normalfont\normalsize
\caption{Mean per-page cumulative token usage over the five repair rounds in Figure~\ref{en:fig:repair-curve}. Persist corresponds to LiveWorkspace with retained Target.}
\label{en:tab:target-token-usage}
\end{table}

Rebuild ends at 91.2 Avg after round five, below LiveWorkspace with retained Target. Retaining Target reduces recorded input and output token usage by \textbf{42.7\%} and \textbf{15.7\%}, respectively. The shared-subset comparison supports Target reuse across repair iterations.

\section{Controlled Editing Protocol}
\label{en:app:editing-protocol}

\subsection{Repeated Instances and Visual Targets}
Each synthetic base page contains \(K\in\{3,4\}\) repeated text elements or cards, with hidden unique \texttt{ItemID} markers used to identify instances during evaluation. CSS decouples source order from visual order using \texttt{flex-direction: column-reverse}, reversed \texttt{order} values, or absolute positioning with reversed vertical offsets. We render the base page and extract Current Shared VS, recording each instance's rendered bounding box, marker, and source index. The target is determined from rendered geometry, for example the bottommost instance or the second instance from the top. Instructions refer to these visual positions rather than source indices. Source order alone therefore does not identify the requested instance.

\subsection{Edits and Task-Specific Verification}
LiveWorkspace updates and saves Target Shared VS according to the user instruction, then passes the updated Target to the same alignment and repair loop used for refinement. After execution, Current Shared VS is recovered for verification. Task-specific checks assess the requested change, while preservation scores assess non-target content.

The cases cover text and color changes, layout changes, and card deletion or reordering. Color edits are verified after rendering the edited page: the instance nearest the recorded target box must have the requested color, and the other repeated instances must retain their colors. Geometric layout checks use rendered Current Shared VS geometry. For deletion and order checks, the scorer parses the edited HTML and compares marker presence and sequence with the recorded ground truth. Deletion requires removal of the target marker while retaining the others; order checks concern marker sequence in HTML. The marker-based checks themselves do not require rendering. All reordering cases additionally undergo independent rendered-page verification to confirm that the final visual order satisfies the instruction. Other editing cases requiring visual inspection are confirmed by manual review.

\subsection{Localized Success and Preservation}
Let \(\mathrm{naive}_i\) indicate whether the target instance completes the requested type of edit, and let \(\mathrm{over}_i\) count modified non-target repeated instances. Edit Success is the mean localized-success indicator,
\[
\mathrm{local}_i=\mathbf{1}[\mathrm{naive}_i=1\ \land\ \mathrm{over}_i=0].
\]
Thus, modifying the correct instance is insufficient if any other repeated instance is also modified. Instance Preservation is
\[
\mathrm{InstPres}_i=1-\frac{\mathrm{over}_i}{K_i-1},
\]
where \(K_i-1\) is the number of non-target repeated instances on the base page. Global Preservation uses rendered Shared VS nodes as its counting unit. The target node is selected among nodes whose text matches the instruction's original label by proximity to the recorded target box. Let \(U_i\) contain all other nodes in the pre-edit rendered state. After before--after node alignment, a node in \(U_i\) is changed if it is unmatched, its normalized text or RGB text color differs, or the geometry/layout checks identify a change. With \(Q_i\subseteq U_i\) denoting these changed nodes,
\[
\mathrm{GlobalPres}_i=1-\frac{|Q_i|}{|U_i|}.
\]
This metric measures preservation of non-target node presence, text, color, and geometry/layout state. Both preservation metrics are computed separately for each case and macro-averaged over the 100 cases. Direct Editing and LiveWorkspace receive the same inputs and matched budget limits; actual usage does not reach the limits.

\section{Repair Metrics}
\label{en:app:repair-metrics}

For Table~\ref{en:tab:repair}, Repair Rate and Collateral use a fixed Target for each page. \textbf{Repair Rate} counts initial discrepancy keys. Each key identifies a Target node and a differing attribute; a missing Target node is represented by a \texttt{MISSING} key. Let \(D_{0,p}\) be the initial discrepancy keys on page \(p\), and let \(R_{p,m}\subseteq D_{0,p}\) be those resolved by condition \(m\). The page-level rate is
\[
\mathrm{RepairRate}_{p,m}=100\frac{|R_{p,m}|}{|D_{0,p}|}.
\]
Reported rates are macro-averaged over pages.

\textbf{Collateral} counts newly defective Target nodes, once per node. Let \(F_{p,0}\) and \(F_{p,m}\) denote the initial and post-repair defective Target-node sets. A Target node is defective if it has at least one attribute difference beyond tolerance or is missing. Then
\[
\mathrm{Collateral}_{p,m}=|F_{p,m}\setminus F_{p,0}|.
\]
The reported value is the mean count per page. This excludes extra Current nodes and additional attribute errors on nodes already defective initially. It is an absolute node count, not a normalized regression rate.

\textbf{Final Fidelity} is evaluated independently of Shared VS Diff using Design2Code's block matching and rendered-page evaluation. Generated and reference implementations are evaluated for Block, Text, Position, Color, and CLIP; their five-score mean is scaled to 0--100. Reference implementations are used by this evaluator, not provided to the reconstruction coder.

  \clearpage
\end{document}